\documentclass[aps,prl,twocolumn,showpacs,superscriptaddress,groupedaddress]{revtex4}
\usepackage{graphicx}  
\usepackage{dcolumn}   
\usepackage{bm}        
\usepackage{amssymb}   
\usepackage{soul}

\usepackage{doi}

\usepackage{tikz}
\usetikzlibrary{arrows}

\usepackage{ulem}
\usepackage{caption}
\usepackage{subcaption}
\usepackage{amsmath}
\newcommand{\YM}[1]{{\color{black}#1}}
\newcommand{\YMs}[1]{{\color{black}#1}}
\newcommand{\YK}[1]{{\color{black}#1}}
\newcommand{\YKs}[1]{{\color{black}#1}}
\newcommand{\e}{\epsilon}
\begin{document}



\title{Abrupt disappearance and reemergence of the \YMs{SU(2)} and SU(4) Kondo effects due to population inversion}
\date{\today}

\author{Yaakov Kleeorin}
\affiliation{Department of Physics, Ben-Gurion University of the
Negev, Beer Sheva 84105, Israel}

\author{Yigal Meir}
\affiliation{Department of Physics, Ben-Gurion University of the
Negev, Beer Sheva 84105, Israel}
\affiliation{The Ilse Katz Institute for Nanoscale Science and Technology, Ben-Gurion University of the
Negev, Beer Sheva 84105, Israel}

\begin{abstract}
 The interplay of almost degenerate levels in quantum dots and molecular junctions with possibly different couplings to the reservoirs has lead to many  observable phenomena, such as the Fano effect, transmission phase slips and the $SU(4)$ Kondo effect.
 \YMs{Here we predict a dramatic repeated disappearance and reemergence of the \YMs{$SU(4)$} and anomalous SU(2)  Kondo effects with increasing gate voltage}. This phenomenon is attributed to the level occupation switching which has been   previously invoked to explain the universal transmission phase slips in the conductance through a quantum dot. We use  analytical arguments and numerical renormalization group calculations to explain the observations and discuss their experimental relevance and dependence on the physical parameters.
\end{abstract}

\maketitle
The coexistence of spin-degenerate levels with different couplings to the leads is ubiquitous in quantum dots (QDs), quantum wires and molecular junctions. It has been pointed out early on that such coexistence may develop in deformed QDs \cite{hackenbroich1997,baltin1999}, with important consequences on the relation between the conductance and the transmission phase of consecutive Coulomb-blockade (CB) peaks. Later on it has been demonstrated \cite{silvestrov2000} that such a coexistence is, in fact, a generic effect in interacting QDs. In fact, the interplay between levels with weak coupling to the leads and a strongly coupled one has been invoked \cite{silvestrov2000,silvestrov2001,golosov2006,kashcheyevs2007,karrasch2007,silvestrov2007} to explain the intriguing experimental observations \cite{yacoby1995,schuster1997,avinun-kalish2005} of  sharp drops in the transmission phase through a QD between CB peaks. These drops have been attributed to "level occupation switching" (LOS) - the abrupt emptying of the strongly coupled level and the filling of a corresponding weakly coupled level, or vice versa, as the gate voltage is continuously varied. These studies have been further supported by a direct observation \cite{aikawa2004} of the Fano effect, resulting from the interference between a wide and a narrow level in a single quantum dot. Simultaneous transport through several molecular levels has also been demonstrated \cite{kiguchi2008} in molecular junctions, and the interplay of weakly and strongly coupled levels has been predicted \cite{bergfield2010} to lead to observable effects in the CB peak structures.

In a seemingly different context, the coexistence of almost degenerate levels has been argued \cite{borda2003,zarand2003,eto2005,choi2005,lopez2005,sato2005,le_hur2007} to give rise to  $SU(4)$ Kondo  \YM{physics} \cite{nozieres1980}, which has indeed been observed in carbon nanotubes \cite{jarillo-herrero2005-2,makarovski2007,anders2008}, in atoms \cite{lansbergen2010} and in single \cite{tettamanzi2012} and double \cite{hubel2008,keller2014} semiconductor quantum dots. In all these systems, the degenerate levels are not necessarily coupled equally to the leads \cite{tosi2015,nishikawa2016}. However, in spite of the plethora of studies of the physics of LOS in QDs on one hand, and of SU(4) Kondo  \YM{physics} in such systems on the other hand, the interplay of these two effects has not been addressed so far. In this letter we predict dramatic abrupt suppression and reentrance of the Kondo effect due to LOS. We present numerical renormalization group (NRG) calculations, backed up by analytical arguments, and show that in the presence of two spin-degenerate levels, with very different couplings to the leads, then as the gate voltage is varied (Fig.~1), the enhanced conductance due to the Kondo effect is abruptly suppressed, only to likewise abruptly reemerge at higher gate voltages. This disappearance and reemergence may occur more than once. Below we elaborate on the physics behind this effect, on its dependence on temperature, the ratio of the couplings of the two levels to the leads, on their energy difference and other physical parameters.

The Hamiltonian that describes the two-level QD is given by
\begin{equation}
H_{QD}=\sum_{i \sigma} \epsilon_{i} \hat{n}_{i \sigma} + \sum_{i} U_i \hat{n}_{i\uparrow}\hat{n}_{i\downarrow} + U_{12} \hat{n}_1 \hat{n}_2
\end{equation}
where $i=1,2$ denotes the level index, $\hat{n}_{i \sigma}=d^\dagger_{i \sigma} d_{i \sigma}$, $\hat{n}_i=\sum_{\sigma} \hat{n}_{i \sigma}$ ($d^\dagger_{i \sigma}$ creates an electron on the dot in level $i$ with spin $\sigma$), and spin-degeneracy has been assumed (i.e. no magnetic field). We will first concentrate on the fourfold degenerate case, $\epsilon_{i}=\epsilon$ and $U_1=U_2=U_{12}=U$. Each one of the levels couples to a different linear combination of states in the leads, which, for simplicity, we assume to be orthogonal. The resulting Hamiltonian is then given by
\begin{equation}
H = H_{QD} + \sum_{i \sigma k\in L,R} \epsilon_{ik}c^\dagger_{i \sigma k} c_{i \sigma k}+ \sum_{i \sigma k\in L,R} \left(V_{ik}d^\dagger_{i \sigma} c_{i \sigma k } + h.c\right) ,
\end{equation}
where $c^\dagger_{i \sigma k}$ creates an electron with spin $\sigma$ in the leads in  the momentum state $k$ that couples to level $i$ in the dot. Again, for simplicity,  the tunneling amplitude is chosen to be momentum (and spin) independent, but different between the two levels,  $V_{i k}=V_i$.
With this separation  the calculation of the linear response current can proceed separately for each channel using the Meir-Wingreen formula \cite{meir1992}, in terms of the spectral function of each level. The spectral function and the expectation values are calculated using a density matrix numerical renormalization group (DM-NRG) procedure \footnote{We used the open-access Budapest Flexible DM-NRG
code, http://www.phy.bme.hu/dmnrg/; O. Legeza,
C. P. Moca, A. I. Toth, I. Weymann, G. Zarand,
arXiv:0809.3143 (2008) (unpublished)}.
 Assuming equal couplings to the left and right leads, and equal density of states $\rho$ in the two leads, both levels only couple to a specific superposition of the left and right lead wavefunctions, and effectively, one needs to solve a single-lead problem per level, characterized by the respective couplings of the two levels, $\Gamma_i=\pi \rho V_i^2$. We assume a constant $\rho$, with a symmetric band around the Fermi energy, with bandwidth $D$. In the following we set $D$ to be the unit of energy.

 \begin{figure}[h]
 	\includegraphics[width=0.5\textwidth,height=0.18\textwidth]{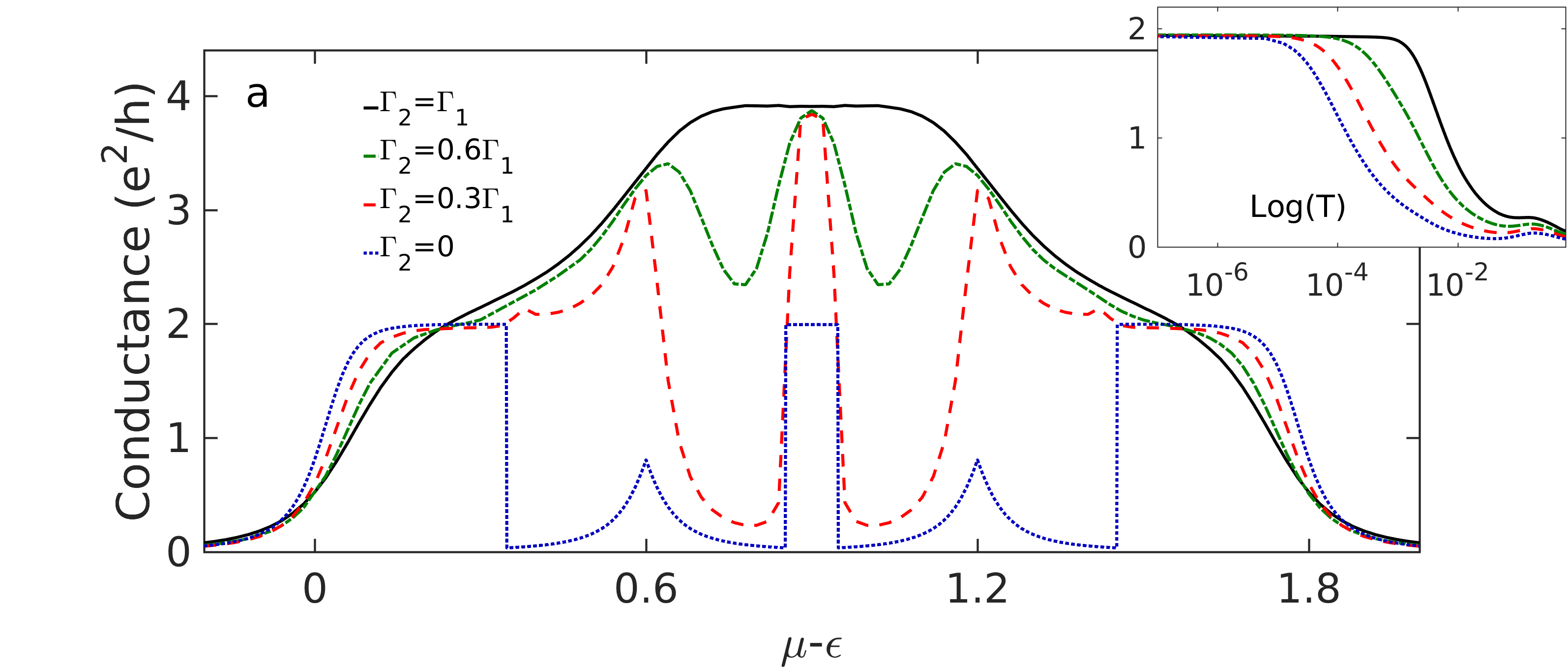}
\includegraphics[width=0.5\textwidth,height=0.18\textwidth]{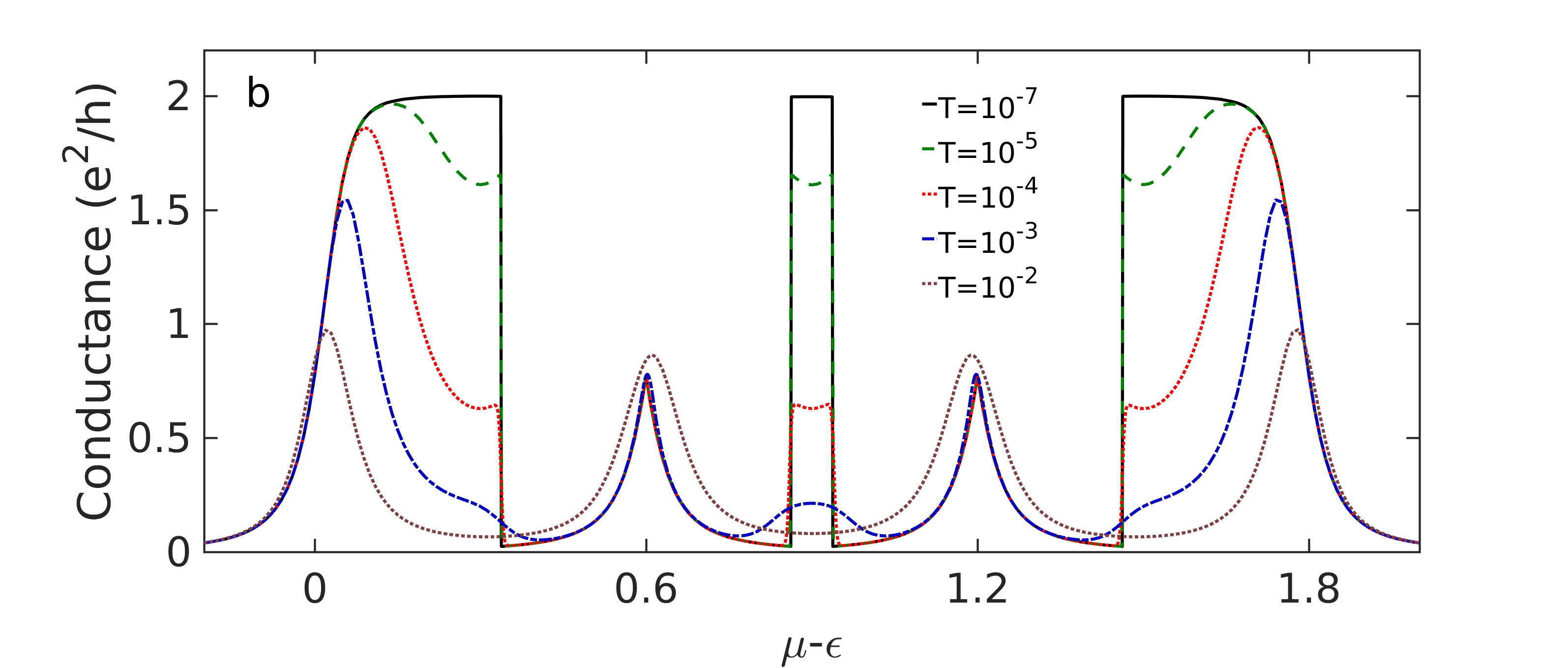}
 	
 	\caption{Conductance through a two-level quantum dot as a function of chemical potential for (a) different ratios of the tunneling amplitudes $\Gamma_2/\Gamma_1$, with $T=10^{-7}$ and (b) different temperatures, for $\Gamma_2=0$. For both plots $\Gamma_1=0.03, U=0.6$ (in units of the band width). In both panels, level population switching events lead to abrupt disappearance and reemergence of the Kondo effect. \YKs{Inset: log-temperature dependence of the conductance for the curves in (a), at $\mu-\epsilon=0.2264$}, depicting the change in the Kondo temperature from its $SU(4)$ value to it $SU(2)$ one.}
 	\label{fig:cond2ch}
 \end{figure}

Fig.~\ref{fig:cond2ch}a depicts the conductance as a function of the chemical potential (gate voltage), for different values of $\Gamma_2/\Gamma_1$. Each $\Gamma_i$ defines an effective $SU(2)$ Kondo temperature $T_K^{(i)}$.
  When $\Gamma_2=\Gamma_1$, one reproduces the standard $SU(4)$-\YM{symmetric Anderson model} conductance plot: the conductance $G$ rises from zero to $G=2e^2/h$ (where $e$ is the electron charge and $h$ the Planck constant), then to $G=4e^2/h$, in agreement with the Friedel sum rule, $G = e^2/h \sum_{i\sigma}\sin^2(\pi n_{i\sigma})$ (where $n_{i\sigma}=<\hat{n}_{i\sigma}>$), which is accurate at such low temperatures. However, when $\Gamma_2$ is reduced, LOS starts to take place, resulting in
  several conductance dips near the mid point. For example, the curve for $\Gamma_2=0.3 \Gamma_1$  exhibits a small peak near the first switching event (at $\mu-\epsilon\simeq 0.35$), and a higher peak near the second switching event (at $\mu-\epsilon\simeq 0.6$), then decreases towards zero, only to abruptly rise again to its unitarity value ($G=4e^2/h$). The effect is even more dramatic for smaller $\Gamma_2$
where $T_K^{(2)} < T$. In this regime, the conductance rises and plateaus at its Kondo value, $G=2 e^2/h$, only to drop sharply to almost zero at a specific value of the chemical potential, slightly above $\mu=\e+U/2$. Then the conductance remains at zero, goes through a narrow peak (around $\mu=\e+U$), of a peculiar shape (see below),  and eventually rises sharply again to around the Kondo value below the mid-point $\mu=\epsilon+3U/2$. Since the model is symmetric around that point, the same behavior is reflected around $\mu=\epsilon+3U/2$.

Fig.~\ref{fig:cond2ch}b depicts how this effect depends on temperature, for the case $\Gamma_2=0$.
 At high temperature, $T>>T_K^{(1)}$, one reproduces the CB peak structure. Note that in spite of only one level being coupled to the leads, there are 4 CB peaks, all of similar width, indicating that in each case transport is through the strongly coupled level \cite{silvestrov2000}. However,   for smaller temperatures, switching events lead to enhancement of the conductance
   by the Kondo effect, but only in specific regions of the chemical potential, giving rise to the sharp drops in the conductance mentioned above.

  This peculiar behavior of the conductance can be understood by combining the Friedel sum rule,
  with the physics of LOS. Fig.~2a depicts the occupations  of the two levels, for $\Gamma_2=0$, at the lowest temperature of Fig.~1b, and the resulting conductance, using the Friedel sum rule for the $\Gamma_2=0$ case: $G = e^2/h \sum_{\sigma}\sin^2(\pi n_{1\sigma})$.

\begin{figure}[h]
	
	\includegraphics[width=0.5\textwidth,height=0.18\textwidth]{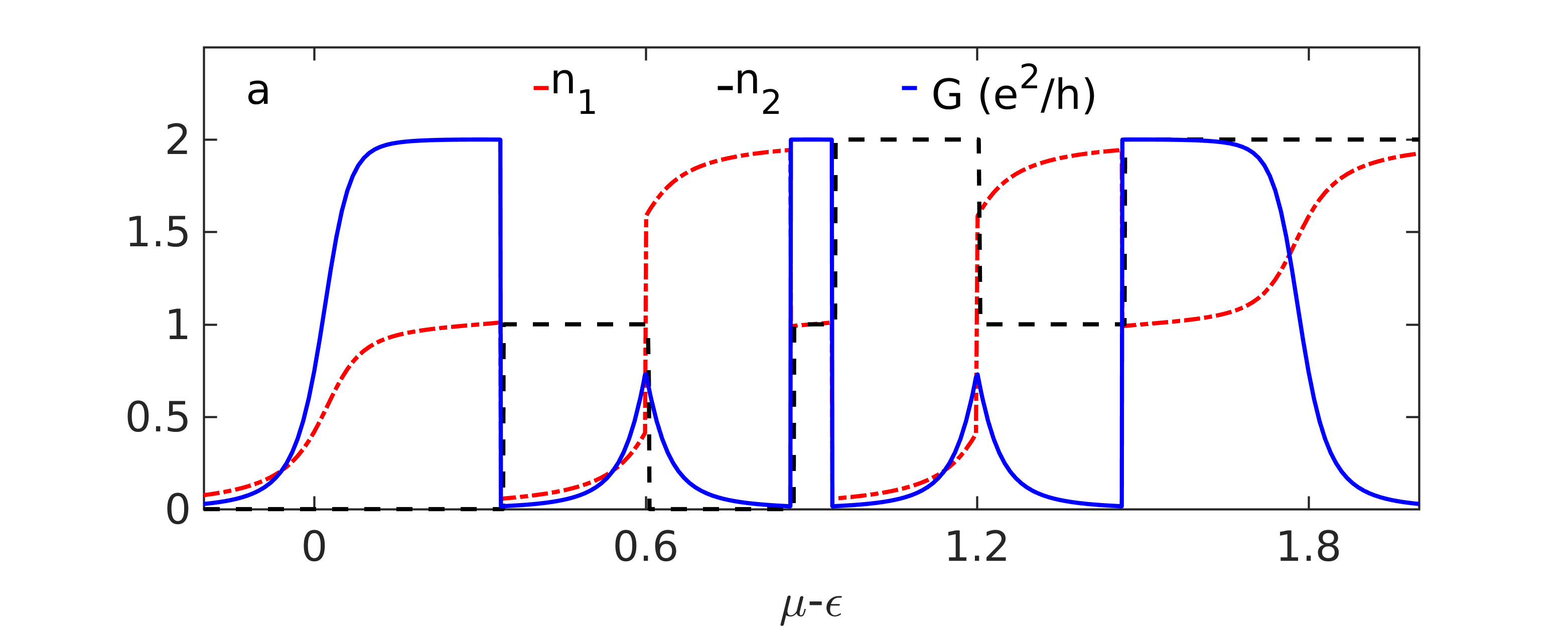}
	\includegraphics[width=0.5\textwidth,height=0.18\textwidth]{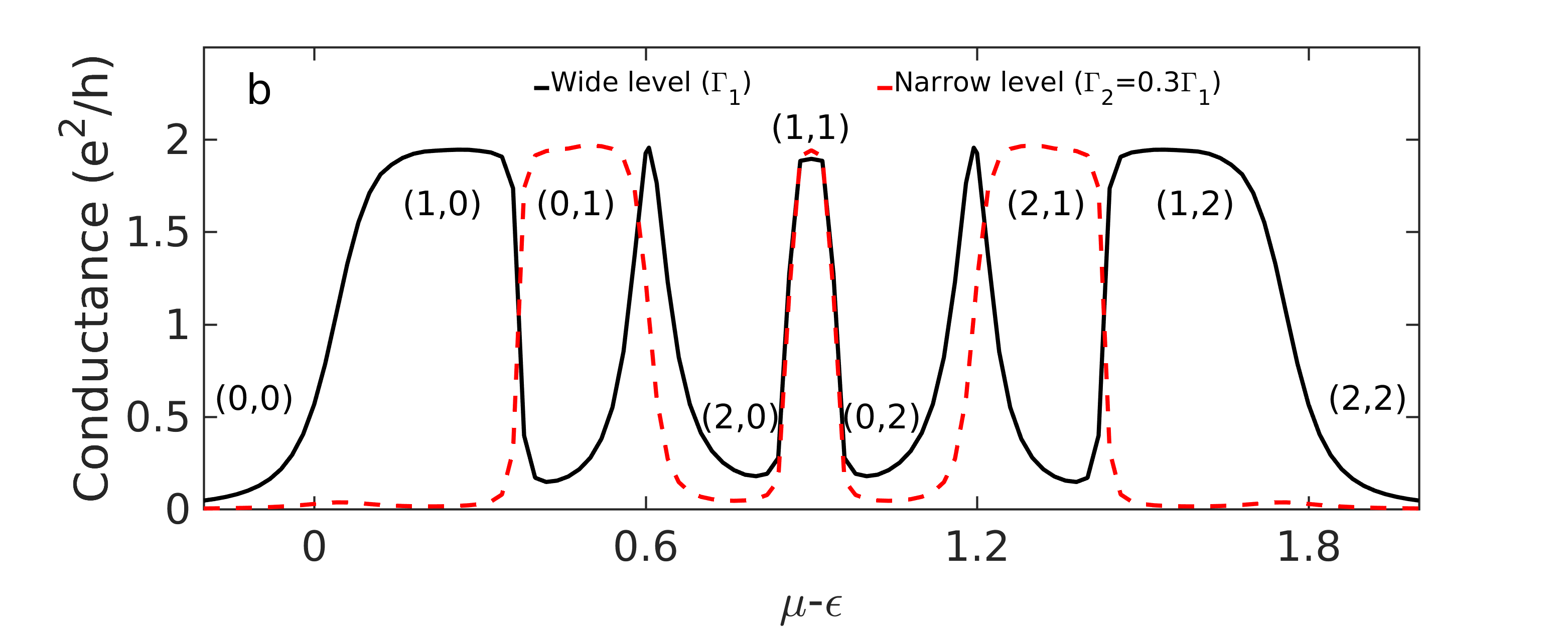}
	\caption{(a) The occupation in each level ($n_1,n_2$) and the conductance $G$, calculated using the Friedel Sum Rule, as a function of chemical potential. Here $\Gamma_1=0.03, \Gamma_2=0, T=10^{-7}$ and $U=0.6$. (b) The separate contributions to the conductance from the wide (with coupling $\Gamma_1$) and the narrow ($\Gamma_2$) levels for $\Gamma_2/\Gamma_1=0.3$, demonstrating a transition between the respective Kondo effects. The occupations for the wide and narrow levels, rounded to a representative integer, are written in parenthesis in each regime.}
	\label{fig:nmu}
\end{figure}

The physics of LOS is relatively well understood \cite{silvestrov2007}. Consider, for example,  the limit of $\Gamma_2=0$. When $\mu$ lies between the first two CB peaks, there is a competition between two configurations: the partially occupied wide level and the fully occupied narrow level. Due to tunneling ($\Gamma_1$), the energy of the former is reduced by an electron process (e-process) -- tunneling of the electron in that level to the leads, and by a hole process (h-process) -- tunneling of an electron from the leads into the dot, making it doubly occupied. On the other hand the second configuration energy is reduced by lead electrons of either spin tunneling into the empty wide level, i.e. twice the h-process. As $\mu$ crosses the symmetry point $\mu=\e+U/2$, the reduction in energy due to the h-process is larger than that of the e-process. As a result of this, it is eventually energetically favorable to occupy the narrow level instead of the wide level, and there is a LOS event. This is depicted in Fig.~\ref{fig:nmu}a, where we plot the occupations of the two levels as a function of chemical potential. We see that similarly to the occupation switching event described above, occurring at $\mu\sim0.33D$ \YK{for the parameters used}, there are several more switching events for similar reasons. As electrical current flows mainly through the strongly coupled level, a sudden switch in its occupation will lead to an abrupt change in the conductance, in accordance with the Friedel sum rule. Indeed, given the occupations, the conductance calculated using the Friedel sum rule (Fig.~\ref{fig:nmu}a) agrees perfectly with the direct calculation of the conductance (Fig.~\ref{fig:cond2ch}).
Similar arguments can be applied to the regime where transport occurs through both levels, e.g. the curve $\Gamma_2=0.3 \Gamma_1$ in Fig.~\ref{fig:cond2ch}, where both Kondo temperatures obey $T_K^{(i)}\gg T$.  In this case, when the occupations switch from around $(n_1,n_2)=(1,0)$ to around $(0,1)$, there is a switch from the Kondo effect due to level 1 to that due to level 2 (Fig.~\ref{fig:nmu}b), resulting in a small hump in the conductance\YM{, visible in  Fig.~\ref{fig:cond2ch}a}. On the other hand, when the occupations switch from $(2,0)$ to $(1,1)$ there is an abrupt jump in the conductance from almost zero to the \YM{coexisting Kondo} value of $G=4e^2/h$. \YKs{The crossover from $SU(4)$ physics to $SU(2)$ physics with decreasing $\Gamma_2/\Gamma_1$ is manifested in the reduction of $T_k$  from $T_k^{SU(4)}$ to $T_k^{SU(2)}$ (inset of Fig.~\ref{fig:cond2ch}a).}

\YK{It is interesting to note the unusual shape of the conductance peak at $\mu=\e+U$ ($\mu\simeq0.6$ in Fig.~\ref{fig:cond2ch}), where the narrow level abruptly empties.} Consider first the case $\Gamma_2=0$. As the chemical potential approaches the value  $\mu=\e+U$, the wide level starts to be gradually filled, its occupation, and as a result, the total conductance, rises as a tail of a Lorentzian of width $\Gamma_1$. At the switching event the occupation of the wide level jumps to almost 2, and the conductance start decreasing, again as a tail of a Lorentzian of width $\Gamma_1$. Thus, the line shape of this peak, in the case of $\Gamma_2=0$, will consist of a cusp formed by the intersection of the tails of two shifted Lorentzians. For a finite $\Gamma_2$, the LOS events will occur on this scale, and we expect an additional narrow Lorentzian of width $\Gamma_2$ on top of the line shape described above.

\begin{figure}[h]
	\includegraphics[width=0.5\textwidth,height=0.18\textwidth]{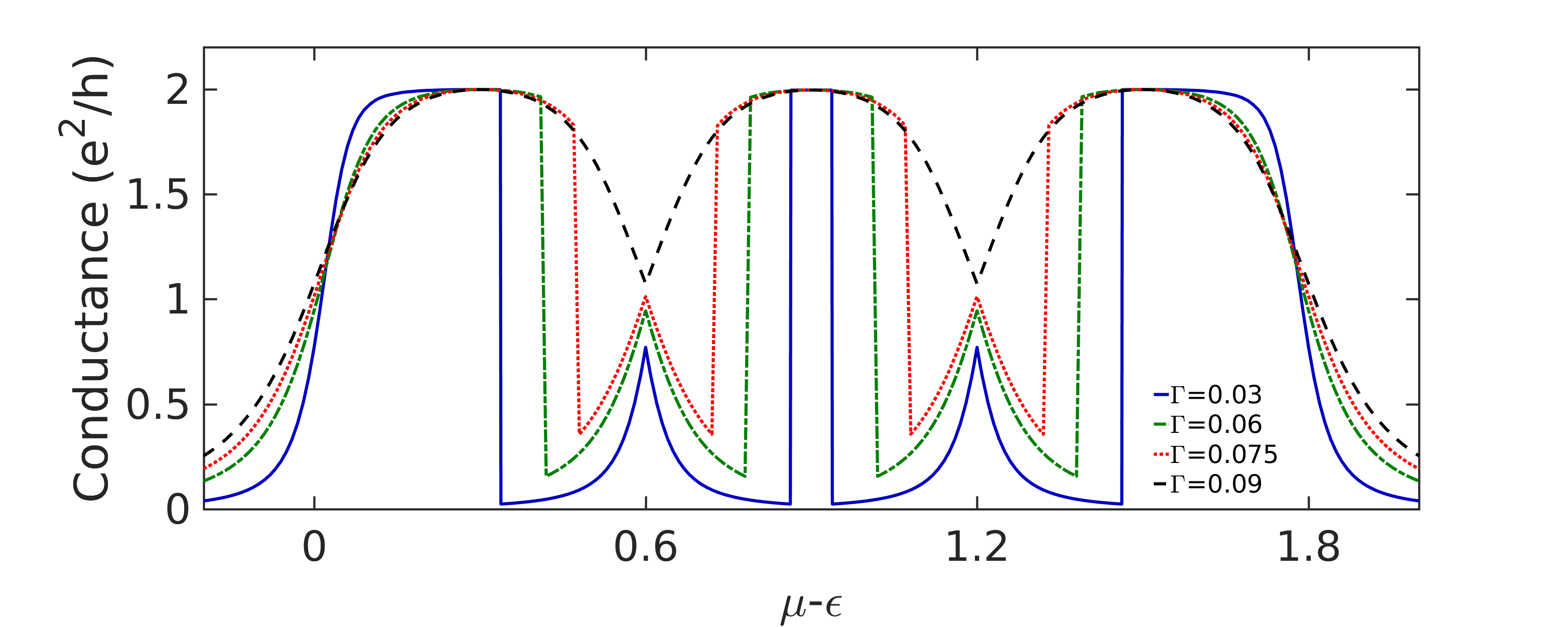}
	\caption{Conductance as a function of chemical potential for various $\Gamma_1$, with $\Gamma_2=0$, $T=10^{-7}$ and $U=0.6$. The larger $\Gamma_1=0.09$ curve exhibit 3 wide Kondo peaks. }
	\label{fig:condG}
\end{figure}

Unlike the perturbative calculation \cite{silvestrov2007}, which predicts the LOS event exactly at the midpoint between the CB peaks, $\mu=\e+U/2$, in the NRG calculation the switching occurs at a higher chemical potential, which shifts to even higher $\mu$ as $\Gamma_1$ increases, as can be seen in Fig~\ref{fig:condG}. In fact, for values of $\Gamma_1$ larger than $\simeq U/10$, the anomalous CB peak at $\mu=\e+U$ turns into a dip. For such large $\Gamma_1$ the LOS events are at $\mu=\e+U$ and $\mu=\e+2U$, where the narrow level becomes occupied by one and two electrons, respectively.
  Between these points the occupation of the wide level rises continuously from $n_1=1/2$  to $n_1=3/2$, and drops sharply back to  $n_1=1/2$  at the switching point.  Thus, in this regime, we find another atypical situation: three consecutive wide Kondo peaks, corresponding to the three possible occupation states of the narrow level.

\begin{figure}[h]
	\includegraphics[width=0.5\textwidth,height=0.2\textwidth]{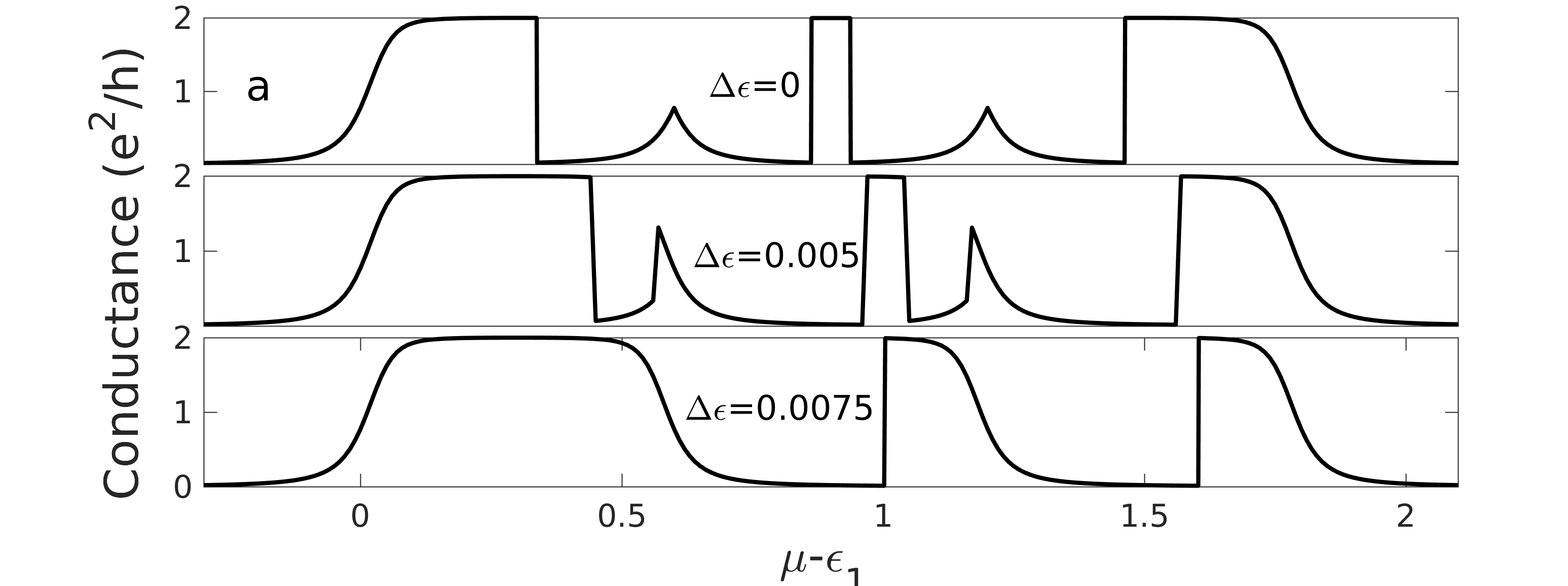}
	\includegraphics[width=0.5\textwidth,height=0.2\textwidth]{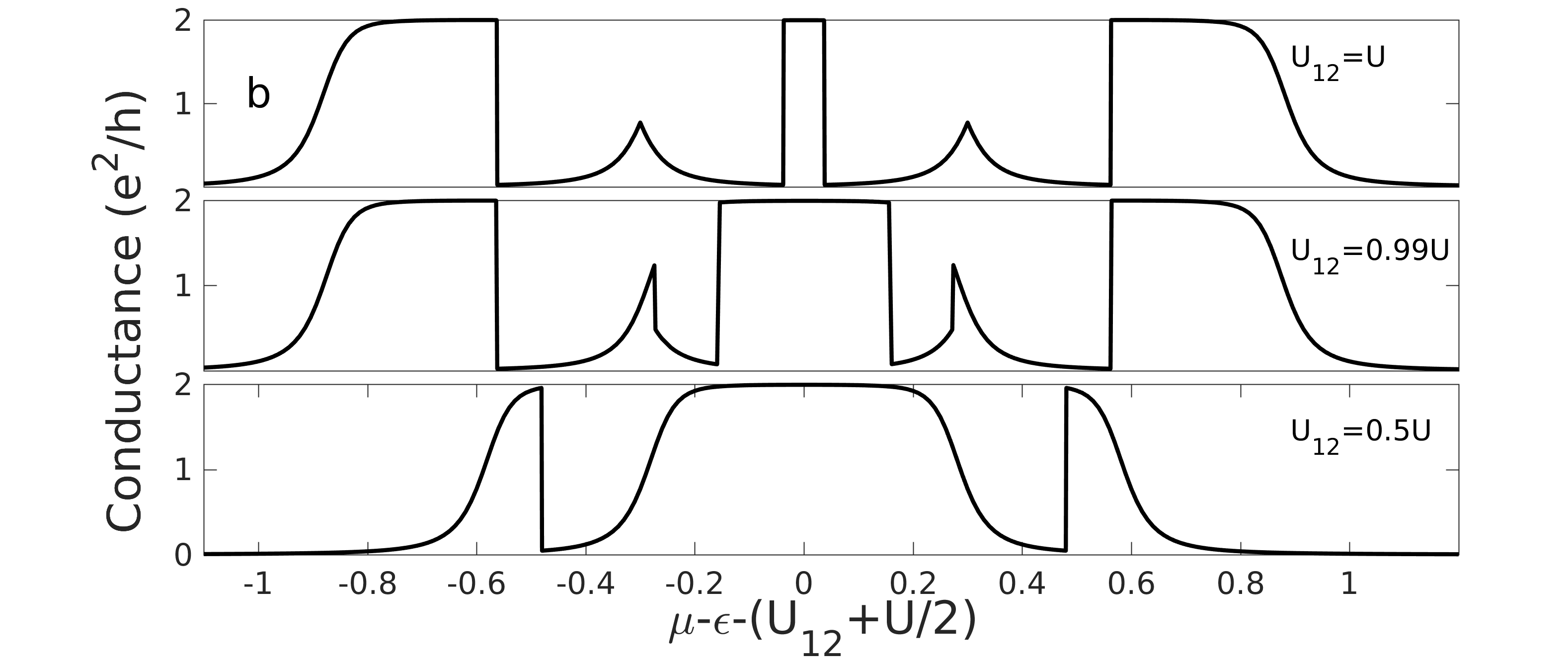}

	\caption  {Conductance as a function of chemical potential for (a) various differences in level energies $\Delta \epsilon=\epsilon_2-\epsilon_1$ and (b) various ratios between inter and intra dot interaction strengths $U_{12}/U$. Curves in (b) are centered around the particle-hole symmetry point $U_{12}+U/2$. Plots were calculated with U=0.6, $\Gamma_1=0.03$ and $T=10^{-7}$.}
	\label{fig:condUe}
\end{figure}

The results we have shown so far were for the fully degenerate case, $\e_1=\e_2$ and $U_{12}=U_1=U_2$. Fig.~\ref{fig:condUe} depicts the conductance as one varies $\Delta\e\equiv\e_2-\e_1$, or $U_{12}/U$ (where $U_1=U_2=U$ were still equal). As $\Delta\e$ increases from zero (Fig.~\ref{fig:condUe}a), the switching point between the first two CB peaks shifts to higher chemical potential,
 until it reaches the second CB peak and disappears\YK{, producing a seemingly standard single-level conductance plot in the Kondo regime.}
  As one expects, at this value of $\Delta\epsilon$, for $\mu\gtrsim\e+U$, as the dot becomes doubly occupied (both electrons occupy the wide level), there should be no Kondo effect. However, as can be seen in Fig.~\ref{fig:condUe}a, there is reentrance into the Kondo regime at higher chemical potential (e.g. $\mu=1$ for $\Delta\e=0.0075$), again to disappear and to reemerge again (e.g. at $\mu=1.6$ for same $\Delta\e$). These reentrances into the Kondo regime occur exactly where the LOS occur: whenever the narrow level gets occupied by an additional electron, the energy of the strongly coupled level shifts up, and its occupation is reduced to below double occupation, leading  to reappearance of the Kondo effect \cite{silvestrov2007}.  When $\Delta\e$ is negative (not shown) one finds the mirror image of the positive-$\Delta\e$ chemical-potential dependence, as now the narrow level is preferentially filled. Thus, even when breaking the energy degeneracy, one still finds  the abrupt transitions and the reentrances that one observes in the degenerate case.

Similarly, when $U_{12}/U$ is reduced from unity the LOS still persists, though the first switching event shifts to lower chemical potential.
 This makes the first plateau (where $n_2=0$) narrower and the regions where $n_2=1$ wider. For smaller $U_{12}/U $, When the second plateau becomes wide enough and the second $n_2=0$ region disappears entirely, the conductance exhibits Kondo peaks that are again abruptly suppressed and then reemerge as the narrow level becomes occupied, with each peak corresponding to a different value of $n_2=0,1,2$.

The observation of the various predictions made in this paper can be checked in different physical setups. One such system would be a single quantum dot, where the physics of level occupation switching has already been demonstrated by the universality in the transmission phase and its abrupt drop between Coulomb-blockade peaks \cite{yacoby1995,schuster1997,avinun-kalish2005}. In such a system, since both levels occupy the same dot, one expects $U_1\simeq U_2 \simeq U_{12}$. Thus if one can reach a regime where the Kondo temperatures associated with the two levels obey $T_K^{(1)}\gg T \gg T_K^{(2)}$, then one should observe the physics described in the this paper. Experimentally, by increasing the coupling of the quantum dot to the leads, the $\pi/2$ phase shift associated with Kondo effect has indeed been observed \cite{ji2002,zaffalon2008}, indicating that $T_K^{(1)}> T $. However, in that regime no abrupt phase drops have been observed, indicating that in this regime the condition  $T > T_K^{(2)}$ has not been met. In order to fulfill this latter condition, one may  either tune the temperature to that regime, or select a narrow level of a smaller $\Gamma_2$. Another relevant example is transport through nanotube quantum dots \cite{nygard2000,jarillo-herrero2005}, where  each orbital level is 4-fold degenerate. Again in this setup one expects, for the same reason, that the Coulomb energies will be of the same order of magnitude. By appropriate application of a magnetic field and gate voltage one can tune the system to have simultaneous transport through two different orbital states, with different coupling to the leads. If the Zeeman splitting is much larger than the coupling of the levels to the leads, then this system will display the  orbital $SU(2)$ Kondo effect, as has been observed in  Ref.\onlinecite{nygard2000}. On the other hand, if the width of the strongly coupled level is larger than the Zeeman splitting, one may observe the abrupt disappearance and reemrgence of the anomalous $SU(4)$ physics detailed in this paper. Another relevant physical system is the  double quantum dot system that was utilized to observe the $SU(4)$ Kondo effect \cite{hubel2008,keller2014}. In this system the two separate quantum dots play the role of the two levels in our theory. Experimentally, one can use gate voltages to tune the QDs energies and couplings to the leads, i.e. the parameters $\Gamma_i$ and $\e_i$. Thus, if one tunes to the already observed $SU(4)$ fixed point, and then gradually reduce the ratio $\Gamma_2/\Gamma_1$, we predict \YKs{a gradual transition from SU(4) Kondo  to SU(2) Kondo behavior and }the eventual emergence of the abrupt suppression and reemergence of the Kondo peak, as detailed in Fig.\ref{fig:cond2ch}. One problem that may be relevant to the two-dot setup is that the inter-dot Coulomb energy $U_{12}$ is typically smaller than the $U_i$, the intra-dot one. In principle, in order to achieve degeneracy, one may tune the difference in energies of the two dots, to compensate for the difference in the Coulomb energy. This physics will be explored elsewhere.

To conclude - we have presented physical arguments and numerical-renormalization-group calculations that demonstrate a dramatic suppression and then reemergence of the $SU(4)/SU(2)$ Kondo effect in quantum dots that contains two spin-degenerate levels, with very different couplings to the leads. Since this has been claimed to be a generic phenomenon in quantum dots and molecular junctions, we expect our results to have a wide range of applicability. In particular, the experiments which have already observed $SU(4)$ Kondo effect, either in carbon nanotube quantum dots, or in semiconductor  quantum dots, could be employed to study the physical regime discussed in this paper, and to critically check our predictions.

We thank P. Moca and G. Zarand for scientific discussions and help with the DM-NRG code. YM acknowledges support from ISF grant 292/15.

\bibliography{mylibrary}

\end{document}